\documentstyle [12pt,epsfig]{article} 
\makeatletter

\@addtoreset{equation}{section}
\makeatother

\newcommand{\ba}{\begin{eqnarray}}
\newcommand{\ea}{\end{eqnarray}}

\begin{document}
\newcommand{\BS}{\bigskip}
\newcommand{\SECTION}[1]{\BS{\large\section{\bf #1}}}
\newcommand{\SUBSECTION}[1]{\BS{\large\subsection{\bf #1}}}
\newcommand{\SUBSUBSECTION}[1]{\BS{\large\subsubsection{\bf #1}}}

\begin{titlepage}
\begin{center}
\vspace*{2cm}
{\large \bf Proposals for Two Satellite-Borne Experiments to Test Relativity
 of Simultaneity in Special Relativity}  
\vspace*{1.5cm}
\end{center}
\begin{center}
{\bf J.H.Field }
\end{center}
\begin{center}
{ 
D\'{e}partement de Physique Nucl\'{e}aire et Corpusculaire
 Universit\'{e} de Gen\`{e}ve . 24, quai Ernest-Ansermet
 CH-1211 Gen\`{e}ve 4.
}
\newline
\newline
   E-mail: john.field@cern.ch
\end{center}
\vspace*{2cm}
\begin{abstract}

An orbiting `photon clock' is proposed to test directly the
 relativity of simultaneity effect of special relativity. This is done by exchanging
 microwave signals between two satellites in low Earth orbit carrying clocks
 that have previously been synchronised in the Earth-centered inertial system.
 A similar experiment using synchronised signals from two GPS
 satellites with the receiver on a single satellite in low Earth orbit, is also
 proposed.

 \par \underline{PACS 03.30.+p}

\vspace*{1cm}
\end{abstract}
\end{titlepage}
 
\SECTION{\bf{Introduction}}
 In Einstein's original 1905 Special Relativity (SR) paper~\cite{Ein1}, three novel 
      space-time phenomena were introduced as physical consquences of the Lorentz transformation, i.e.
     as purely kinematical effects. These are: Relativity of Simultaneity (RS)\footnote{
      In fact, the RS effect discussed in Ref.~\cite{Ein1} is not the one described
     below in Section 3, that is correlated with the LC effect, but follows from a discussion
     of light signal clock synchronisation in different inertial frames, without reference to
     the Lorentz transformation. The RS effect tested by the proposed experiments is 
     the standard text-book one derived, like LC, from the Lorentz transformation.} 
     relativistic Length Contraction (LC) and Time Dilation (TD). Many direct and
   precise experiments have confirmed the existence of TD; however to date, there is no
   experimental confirmation of either RS or LC. It is the purpose of the present
   article to propose definitive experimental tests of  RS.
 \par Unlike essentially all recent high precision texts of special relativity,
       the aim of the proposed experiments is not to search for a breakdown
     of SR but rather to test a particular prediction of standard SR.  This important
    point was not understood by some reviewers of earlier versions of this paper
    who stated, incorrectly, that existing experimental limits on SR test parameters
     (for example those of the Mansouri and Sexl test theory~\cite{MS}) show that SR is
    experimentally confirmed, so therefore there is `nothing to test'. For this 
     reason, in the following section, a `mini-review' of current experimental
     tests is presented, where the complete lack of overlap between conclusions
     drawn from the results of previous experiments, and those that may be obtained from the
     proposed experiments, is pointed out. The standard relativity of simultaneity
    effect of SR is explained in Section 3.
    The following sections present the proposed experiments and a comparative discussion
    of the RS test proposed by F\"{u}rth~\cite{Furth} in 1965. A brief summary 
     is provided in the concluding section. 

\SECTION{\bf{Mini-review of experimental tests of special relativity}}

  The availability of precise atomic clocks and frequency stabilised lasers has enabled an enormous
 increase, over the last four decades, in the precision of of some experimental tests of
 SR. In particular, much improved experimental upper limits on the
 anisotropy of the speed of light, testing Local Lorentz Invariance (LLI), have
 accompanied the development of sophisticated theoretical models such as the 
 `standard model extension'~\cite{SME} as well as more specific models employing
 field-~\cite{FT} or string-~\cite{ST} theoretical ideas. The most widely used
 test theory for SR and LLI is, however, the kinematical one of Robertson~\cite{Rob}
 and Mansouri and Sexl~\cite{MS}(RMS).
  This model assumes the existence of a 
  `preferred frame', $\Sigma$, (often identified with the proper frame of the cosmic
   microwave background) relative to which an arbitary inertial frame, S, is in 
   motion with velocity $\vec{v} = c \vec{\beta}$. Denoting time and spatial position
   in $\Sigma$ as ($T$;$\vec{X}$) and in S as ($t$;$\vec{x}$), and assuming conventional
   Einstein light-signal clock synchronisation, the model postulates the following space time
   transformation equations: 
\begin{eqnarray}
    x & = & b(X-vT), \\
    t & = & (a+b\beta^2)T-\frac{b \beta X}{c}, \\
     y& = &Yd,~~~z = Zd
\end{eqnarray}
 where, in SR,
 \begin{equation}
  b = \frac{1}{a} = \gamma = \frac{1}{\sqrt{1-\beta^2}},~~~d = 1. 
\end{equation}
 in which case (2.1)-(2.3) become the Lorentz transformation (LT) and $\Sigma$ 
 another inertial frame. Since in most tests of SR and LLI the leading relativistic
 terms are of O($\beta^2$), it is convenient to define further parameters 
  $\tilde{\alpha}$,  $\tilde{\beta}$ and  $\tilde{\delta}$, such that:
 \begin{equation}
  a = 1 +\tilde{\alpha}\beta^2+...,~~~ b = 1 +\tilde{\beta}\beta^2+...,~~~
  d = 1 +\tilde{\delta}\beta^2+...~.
\end{equation}
 In SR $-\tilde{\alpha} = \tilde{\beta} = 1/2$, $\tilde{\delta} = 0$. 
  \par The most precise tests of the isotropy of the speed of light are modern versions
   of the Michelson-Morley~\cite{MM}(MM) and Kennedy-Thorndike~\cite{KT}(KT) experiments.
  The MM experiment tests the angular dependence of light speed anisotropy. A recent
  limit~\cite{HSKNP}, expressed in terms the parameters of the RMS test model is:
    \[\tilde{\beta}- \tilde{\delta}-\frac{1}{2} = (-2.1\pm 1.9) \times 10^{-10} \]
    The KT experiments are sensitive to the velocity dependence of light speed anisotropy,
     recent results~\cite{WBCLST} giving:
    \[\tilde{\beta}- \tilde{\alpha} - 1 = (-3.1\pm 6.9) \times 10^{-7} \]
  The parameter  $\tilde{\alpha}$ is measured in experiments (successors of that of
   Ives and Stillwell~\cite{IS}) on time dilation or the transverse Doppler shift.
    An experiment using collinear saturation spectroscopy of optical transitions
   in a beam of $^7{\rm Li}^+$ with $\beta = 0.064$~\cite{SKEHK} obtains the limit:
    \[ \tilde{\alpha}  +\frac{1}{2} < 2.2 \times 10^{-7} \]
    Another experiment sensitive to $\tilde{\alpha}$, of particular interest, in view of
   the second experiment proposed in the present letter, uses the GPS satellite system to test
   for anisotropy of the one-way speed of light~\cite{WP}. It was found that 
    $\delta c/c < 2 \times 10^{-9}$ corresponding to the limit:
       \[ |\tilde{\alpha}  +\frac{1}{2}| < 10^{-6} \]
    \par The RMS test model parameter $a$ is called the `time dilation' parameter.
    That this nomenclature is correct is evident from Eqns(2.1) and (2.2). Suppose
    a clock is at rest at $x =0$ in S. Eqn(2.1) shows that, in $\Sigma$, it has the space-time
    trajectory $X = v T$. Substituting this relation in (2.2) gives immediately
   $t = a T$, which becomes, in SR,  $T = \gamma t$ which is indeed the TD effect.
    The parameter $b$ is commonly called the `length contraction parameter'
    which looks plausible in view of Eqn(2.1). The LC effect is however more subtle than
    simple inspection of Eqn(2.1) might suggest. Two space-time events and two clocks
    are involved, and the synchronisation of distant clocks is crucial. This means that
   the limit on $\tilde{\beta}$ provided by the MM or KT experiments does not confirm or 
   invalidate the existence of LC. The relation between the RS and LC effects~\cite{JHF2}
   and the question
   of their existence~\cite{JHF3} have been previously discussed by the present author.
    Reference~\cite{JHF3} also contains a concise review of the experimental tests, to date, of SR.
   In any case, if it does exist, LC is certainly
   an O($\beta^2$) effect, and the experiments proposed here, to test for RS, 
     are quite insensitive to it.

\SECTION{\bf{The relativity of simultaneity effect}}
   \par To discuss RS it is convenient to use what Mansouri and Sexl call `system external
    synchronisation'~\cite{MS}. Suppose that there are clocks at $x= 0$ and $x=l$
    in the frame S. Two clocks at rest in $\Sigma$, spatially contiguous with the
   clocks in S at some instant, are simultaneously synchronised with those in S. That is 
    both clocks in $\Sigma$ are set to $T=0$ and both clocks in S to $t =0$. The
    transformation describing the relation
    between the positions and times in  $\Sigma$ and S of the clock at $x = 0$ is then given,
    in the RMS test model, by Eqns(2.1) and (2.2) with $x = 0$, whereas the corresponding equations for the
    clock at $x = l$ are instead:
\begin{eqnarray}
    x-l & = & b(X-L-vT), \\
    t & = & (a+b\beta^2)T-\frac{b \beta(X-L)}{c}.
\end{eqnarray}
    Substituting $x = l$, $t = T = 0$ in (3.1) and (3.2) shows that these equations are verified
   providing that $X = L$ at this instant. At the later time $T$, in the frame $\Sigma$,
   Eqns(2.1) and (2.2) give, for the
   time of the clock in S at $x=0$, $t(x=0) = aT$,
  while (3.1) and (3.2) give for the
  time of the clock at $x = l$ in S, $t(x=l) = aT$. Therefore, in SR:
   \begin{equation}
    t(x=0) = t(x=l) = \frac{T}{\gamma}.
  \end{equation}
   Both clocks show the same TD effect, but there is no RS\footnote{The same physical effect
    has been considered by Mansouri and Sexl as the consequence of a specfic synchronisation
    procedure of the clocks in the frame S. See Eqns(3.6) of the first reference in
    ~\cite{MS}.}.
     If now the substitution $x = l$ is made in (2.2), instead of 
     (3.2), the relation between $T$ and $t(x=l)$ and $T$ is found to be:
     \begin{equation}
   t(x=l) = (a+b\beta^2)T - \frac{b \beta X(x= l,T) L}{c}.  
  \end{equation}   
   So that instead of (3.3) it is found in SR that
    \begin{equation}
    t(x=0) = t(x=l) +\frac{\gamma \beta L}{c} = \frac{T}{\gamma}  
  \end{equation}
    where $L \equiv  X(x= l,T)-X(x= 0,T)$  
  This means that, at any epoch, $T$, in the frame  $\Sigma$, the clocks in S at $x= 0$ and $x=l$ show
   times differing by $\gamma \beta L/c$. That is, events that
  are simultaneous in $\Sigma$ are not so, according to the clocks in S. Eqn(3.5) corresponds to
    the standard text-book presentation of RS~\cite{TL}. Note that both
   (3.3) and (3.5) are predictions of SR. They differ only in the manner in which the spatial
   coordinates of the  clock at $x =l$ are inserted into the Lorentz transformation equations.
    The possible correctness of (3.3) does not therefore mean that
    $a \ne 1/\gamma$ or  $b \ne \gamma$ in (3.4). In both cases $a = 1/b =1/\gamma$.
    Eqns(3.3) and (3.5) differ, not in the parameters of the test model, but by a different description
   of initial conditions
   in the LT equations, without any violation of LLI. The two experiments proposed below
   can clearly distinguish between the two possibilities (3.3) (no RS) and (3.5) (RS exists).
   It is important for this that the RS time difference in (3.5), $\gamma\beta L/c$, is an
   O($\beta$) not an O($\beta^2$) effect.

 \SECTION{\bf{An Einstein light-signal clock in space}}

 \begin{figure}[htbp]
\begin{center}\hspace*{-0.5cm}\mbox{
\epsfysize15.0cm\epsffile{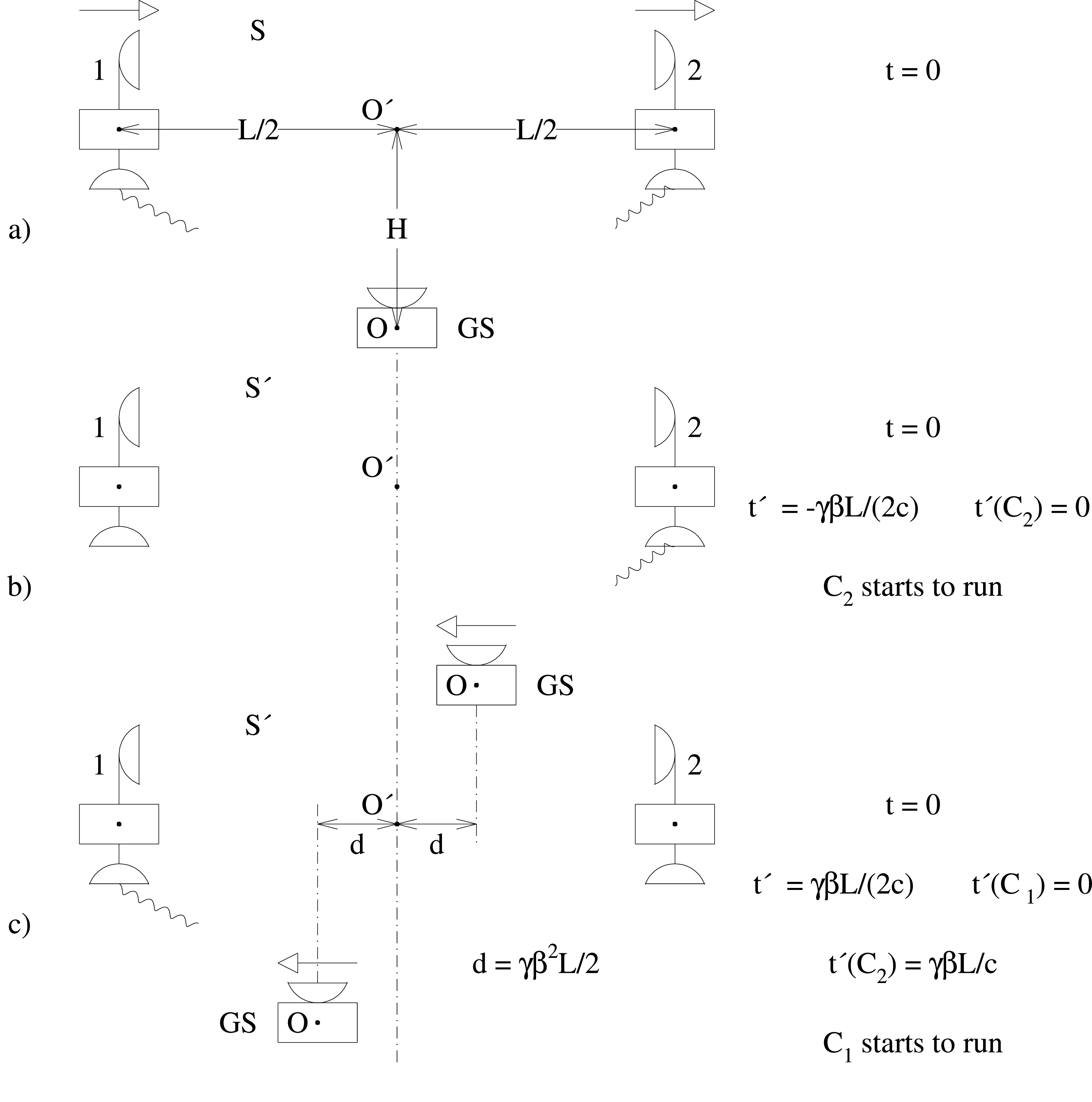}}
\caption{{\em `System external' synchronisation of clocks on two satellites in low-Earth orbit.
 a) Simultaneous microwave signals are sent from the ground station GS so as to arrive 
  simultaneously in the ECI frame at Satellites 1 and 2. b) Signal arrives at Satellite 2
 in the comoving inertial frame S' of the satellites and starts clock C$_2$. c)
  Signal arrives at Satellite 1 in S' and starts clock C$_1$  The configurations in b) and c)
  are calculated using Eqs.~(2.1) and (2.2) with $b = 1/a = \gamma$, $d = 1$. See text for further
  discussion.}}
\label{fig-fig1}
\end{center}
\end{figure}

\begin{figure}[htbp]
\begin{center}\hspace*{-0.5cm}\mbox{
\epsfysize12.0cm\epsffile{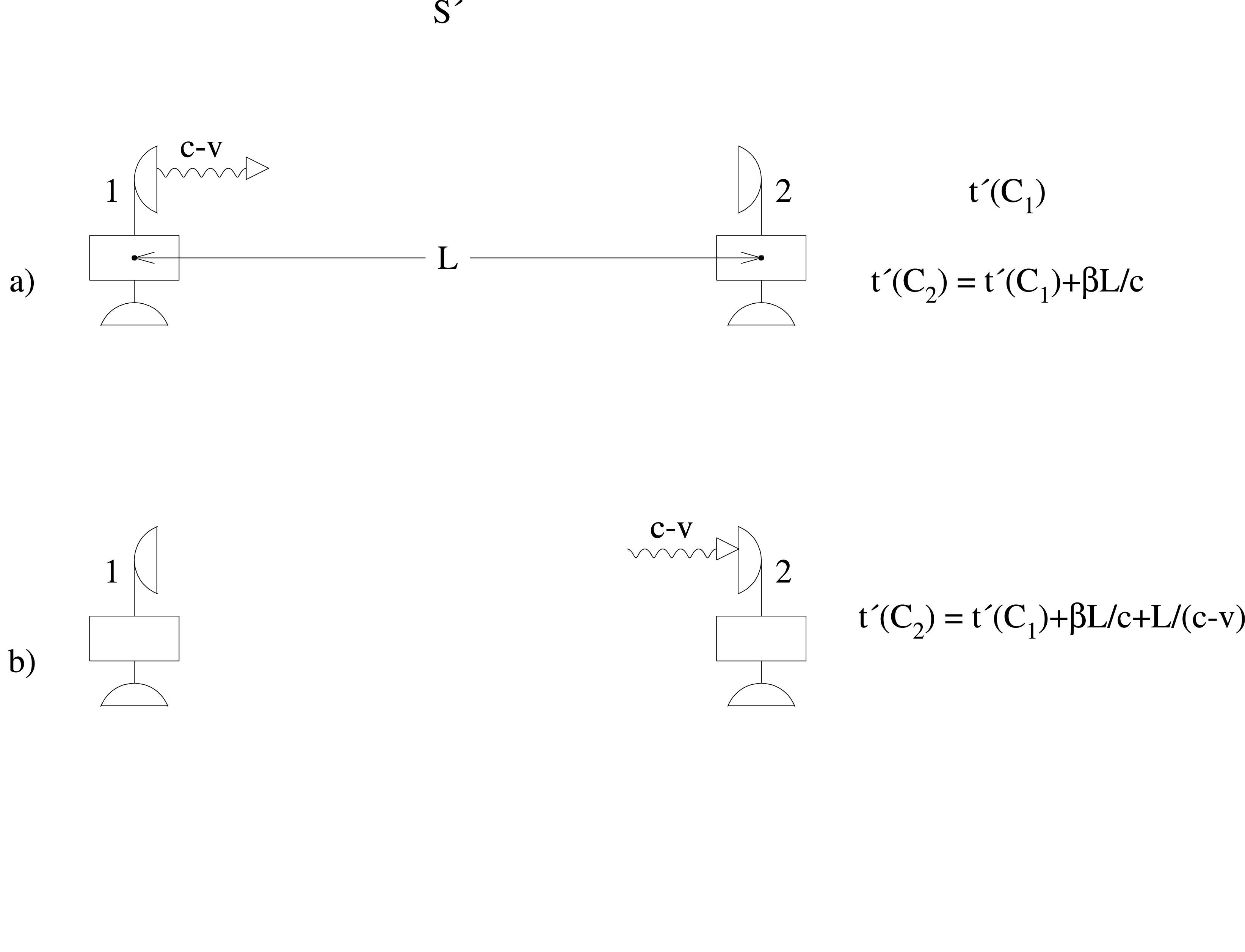}}
\caption{{\em Exchange of a microwave signal between the previously synchronised Satellites 1 and 2
  as viewed in the common comoving inertial frame S' of the satellites. a) The signal is sent at epoch
   $t'({\rm C}_1)$ from Satellite 1 and received [b)] at epoch  $t'({\rm C}_2)$ by Satellite 2.
    O($\beta^2$) corrections are neglected. See text for further discussion.}}
\label{fig-fig2}
\end{center}
\end{figure}

    \par The first experiment proposed makes use of the light-signal clock synchronisation procedure described
     by Einstein in ~\cite{Ein1}. The proposed experiment requires two satellites, one of which could
    conveniently be the International Space Station (ISS) the other a satellite,
   following the same orbit as the ISS, but separated from it by a few hundred 
   kilometers.

   \par A scheme of the proposed experiment is shown in Fig.1. The satellites, 1 and 2, in
    low Earth orbit, separated by the distance $L$, pass near to a ground station GS. Cartesian
    coordinate systems are defined in the co-moving inertial frame of the satellites (S')
   and the ground station (S). The origin of S' is chosen midway between 1 and 2 with
   $x'$-axis parallel to the direction of motion of the satellites and $y'$  axis outwardly
   directed in the plane of the orbit. O$x$ and O$y$ are parallel to  O$x'$ and O$y'$
   at the position of closest approach (culmination) of O' to GS  At culmination, the coordinates of O' in S are:
   (x,y,z) = (0,$H$,$D$) and the relative velocity of frames S and S' is  parallel to the $x$-axis
    of magnitude $v = c \beta$\footnote{It is assumed in the calculations that the change of the co-moving inertial frame due
    to the actual, nearly circular, orbital motion of the satellites may be neglected for small values of
     $t$ and $t'$. With an inter-satellite distance of 400 km the satellites
    move about 11 m during the transit time of the signal between them. This corresponds to a change in the direction
   of the comoving system of 1.6 $\mu$rad and a lateral shift of the reciever position of only 17 $\mu$m. }. As shown in Fig.~1a, clocks C$_1$, C$_2$ in the satellites 1, 2 are synchronised in the frame S
      at time $t = 0$ by reception of simultaneous microwave signals sent from GS at such a time as to arrive simultaneously
      at the satellites when O and O' are aligned. The corresponding times of arrival in the frame S' as predicted by the
       usual Lorentz transformation equations ((2.1) and (2.2) with $x = x'$, $t = t'$, $X = x$, $T=t$, 
       $b=1/a = \gamma$ and $d = 1$) are $t' = -\gamma\beta L/(2c)$ for Satellite 1 and $t' = \gamma\beta L/(2c)$
        for Satellite 2. On reception of the signals, the identical clocks C$_1$ and C$_2$, which are previously stopped and set to
        record an epoch zero, are started. After reception of the signal by Satellite 1 the epochs recorded by the
       clocks at any instant in the frame S' therefore satisfy the relation:
       \begin{equation}
       t'({\rm C}_2) = t'({\rm C}_1)+ \frac{\gamma \beta L}{c}.
       \end{equation} 
        The configurations of the clocks and the ground station, as viewed in the frame S', on 
       reception of the synchronistion signals by Satellites 2 and 1 respectively, are shown in Fig.~1b and 1c, 
       together with the epochs registered by the running clocks. 
     \par In order to test the RS effect that is manifest in Eq.~(4.1) it is sufficient to send, as in Einstein's 
      clock synchronisation procedure, a light signal from from  Satellite 1 to Satellite 2 (or vice versa) at
      a known epoch of the clock on the signal-emitting satellite and observe its time of reception 
      at the other one. As in the analysis of the operation of the GPS system~\cite{AshbyPT,Aetal} it is important 
     to take properly include the Sagnac effect~\cite{Sagnac,Post} in the calculation of the flight
     time of the signal. This is done by taking into account the relative velocity of the light signal,
     which is assumed to move at the universal speed $c$ in the Earth-Centered (ECI) frame, and the
     signal receiver. The ECI frame is instantaeously comoving with the centroid of the Earth, with origin
     at the latter, and has coordinate axes with constant orientations relative to the fixed stars. It is the frame
    in which the synchronisation of the clocks in the satellites of the GPS system is defined~\cite{AshbyPT}.
     Since the speed of the ground station in the ECI frame due to the rotation of the Earth ($< 0.5$km/s)
    is considerably less than that of a low-Earth-orbit satellite ($7.7$km/s for the International
   Space Station (ISS)~\cite{ISS}) 
   the relative velocity, $v$, of the frames S and S' is, to a good approximation, equal to that of the satellite
    relative to the ECI frame that is needed to calculate the Sagnac effect. 
    \par The exchange of a light signal between Satellite 1 and Satellite 2 as viewed in their common comoving
     frame S' is shown in Fig.~2. If the signal is emitted at time $t'({\rm C}_1) = \tilde{t}'_1$ the predicted
       time of arrival as recorded by the clock ${\rm C}_2$ is given by (4.1) and the Sagnac effect as
      $t'({\rm C}_2) = \tilde{t}'_2$ where 
           \begin{equation} 
       \Delta t'_{{\rm RS}}\equiv \tilde{t}'_2- \tilde{t}'_1 = \frac{L'}{c-v} +\frac{\beta L'}{c} = 
          \frac{L}{c} + \frac{2\beta L}{c}+ {\rm O}(\beta^2)~~~~~~({\rm RS})
        \end{equation} 
      where $L' \equiv \gamma L$.  In the absence of the RS effect (i.e. employing Eqs.~(3.1) and (3.2) to give the
       SR prediction  instead of Eqs.~(2.1) and (2.2)) it is found that
          \begin{equation} 
       \Delta t'_{{\rm No~ RS}}\equiv \tilde{t}'_2- \tilde{t}'_1 = \frac{L}{c-v} = 
          \frac{L}{c} + \frac{\beta L}{c}+ {\rm O}(\beta^2).~~~~~~({\rm No~RS})
        \end{equation}     
        Comparing (4.2) and (4.3) it is seen that
            \begin{equation} 
           \Delta t'_{{\rm RS}}- \Delta t'_{{\rm No~ RS}} =  \frac{\beta L}{c}.
         \end{equation}     
         \par If the satellites have the orbital velocity of the ISS, $v = 7.67$ km/s ($\beta = 2.56\times 10^{-5}$)
            and choosing $L = 400$ km, the time-of-flight of the microwave signal between the satellites is
            $\simeq 1.4$ ms and $\beta L/c = 34.2$ ns to be compared with a time resolution of better than
            1 ns in, for example, the NAVEX low Earth orbit satellite experiment~\cite{NAVEX}. The actual distance $\cal L$ separating 
       the satellites during the measurement of $\Delta t'$ can conveniently be found by measuring the
        time of arrival at Satellite 1,  $t'({\rm C}_1) =\tilde{t}'_{1R}$, of the signal reflected back from 
       Satellite 2.
          Then
     \begin{equation} 
         \tilde{t}'_{1R}- \tilde{t}'_{1}  =  \frac{{\cal L }}{c-v}+ \frac{{\cal L }}{c+v}
     =  \frac{{\cal L }}{c(1-\beta^2)}
   \end{equation}  
       so that 
       \begin{equation} 
          {\cal L} = \frac{c(1-\beta^2)( \tilde{t}'_{1R}- \tilde{t}'_{1})}{2}
        = \frac{c(\tilde{t}'_{1R}- \tilde{t}'_{1})}{2}+ {\rm O}(\beta^2)
   \end{equation} 
     and
     \begin{eqnarray}
       \Delta t'_{{\rm RS}} & = & \frac{(1+2 \beta)}{2}( \tilde{t}'_{1R}- \tilde{t}'_{1})+ {\rm O}(\beta^2), \\
    \Delta t'_{{\rm No~ RS}} & = & \frac{(1+ \beta)}{2}( \tilde{t}'_{1R}- \tilde{t}'_{1})+ {\rm O}(\beta^2).
    \end{eqnarray}
   \par The ease of measurement of the O($\beta$) RS effect, evident from the predicted results
      of the experiment just described,  may be contrasted
   with the difficulty of measuring, in a similar experiment, the  O($\beta^2$)
   LC effect. Using the value  $\beta = 2.56\times 10^{-5}$ appropriate for the ISS,
  the apparent contraction of the distance between the satellites 1 and 2,
  as viewed at some instant in S, of $(1-1/\gamma)L$, amounts to only 131$ \mu$m for
   $L = 400$ km. It is hard to see how any
   experiment using currently known techniques could have a sufficiently good spatial resolution
     to measure such a tiny effect.

 \SECTION{\bf{Comparison with the experiment proposed by F\"{u}rth to detect RS}}
 In this section the experiment described above is compared with that proposed by F\"{u}rth
 in 1965~\cite{Furth}. In the latter, the phase difference between signals from
  two spatially-separated phase-synchronised microwave transmitters as measured by a 
  receiver, either at rest or in motion relative to the transmitters, was considered. Since the phase
  of an electromagnetic wave is unchanged by propagation in free space, the phase difference
  $\Delta \phi_0$ measured by the stationary receiver is given by the time-of-flight of
  the signal from the further transmitter to the nearer one: $\Delta \phi_0 = 2 \pi \nu l/c$.  Here $l$ is the separation
  of the transmitters parallel to the direction of observation and $\nu$ is the microwave
  frequency. If the receiver is in uniform motion, parallel to the direction of the microwave signals,
  with speed $v$ relative to the transmitters, use of the conventional LT predicts, as in Eq.~(3.5), an additional phase difference due
  to the RS effect:
   \begin{equation}
    \Delta \phi_v =  \Delta \phi_0 + \frac{2 \pi \nu l v}{c}~~~(c' = c-v) 
   \end{equation}
    where only the lowest order, O($\beta$), velocity-dependent term is retained, and the relative speed $c'$ of
   the signals and the receiver, in the rest frame of the latter, due to the Sagnac effect, is $c' = c-v$.
    \par It was argued by Berry {\it et al}~\cite{Betal} that the phase shift $\Delta \phi_0$ in the above equation,
     corresponding to the passage of the signal between the two transmitters would be modified, at O($\beta$), in 
    the rest frame of the moving receiver. Assuming that the light signals move at the same
    speed $c$ in the rest frames of both the transmitters and the receiver, as predicted by the
    conventional  Relativistic Parallel Velocity Addition Relation (RPVAR):\newline $w = (u+v)/(1+uv/c^2)$ of SR,
     the relative velocity
    between the signals and the transmitters in the rest frame of the receiver is found to be $c+v$
    This gives a different prediction for $ \Delta \phi_v$: 
  \begin{equation}
    \Delta \phi_v = \frac{2 \pi \nu l}{c+v} + \frac{2 \pi \nu l v}{c} = \Delta \phi_0 + {\rm O}(\beta^2).~~~(c' = c) 
   \end{equation}
  At O($\beta$) the phase shift due to RS is cancelled by a term arising from the different signal 
  flight time in the receiver rest frame. This calculation~\cite{Betal} does not however take correctly
  into account the lowest order Sagnac effect. At O($\beta$) the relative velocity of the signals and
   the transmitters is {\it the same} in the rest frame of the transmitters and that of the receiver. If this were
   not the case the experimentally confirmed Sagnac effect simply would not occur. If then the experiment
   proposed by  F\"{u}rth were to be performed, and the same phase shift measured for
   stationary and moving receivers, this is evidence that the RS effect does not exist, not as claimed in Ref.~\cite{Betal}
   that this is the correct prediction of standard SR with RS and the conventional RPVAR.
   If the latter were applicable in the experiment under consideration then the Sagnac effect would not occur.
   \par There is a close similarity between the experiment proposed in the previous section of the
     present paper and that
       proposed by F\"{u}rth. In the former the clocks are synchronised in a frame in which they are in uniform 
      motion and then later compared in their common comoving frame. In the latter, microwave transmitters
     (effectively clocks) are synchronised in their common rest frame and then compared in a frame in  
      which they are in motion. In both cases time (or phase) differences are generated by signal
     propagation between the clocks (or transmitters). In both experiments the relative speed (at lowest order in $v$)
    of the signals and their receiver is $c-v$. Setting (incorrectly) the relative speed of the signals and
     the receiver to $c$, in the rest frame of the receiver, in either experiment, yields similar predictions. Instead of (4.2) it is found that
   \begin{equation} 
       \Delta t'_{{\rm RS}} = \frac{L}{c} +\frac{\beta L}{c}+ {\rm O}(\beta^2)~~~(c' = c) 
        \end{equation} 
    and instead of (4.3)
          \begin{equation} 
       \Delta t'_{{\rm No~ RS}} = \frac{L}{c}+ {\rm O}(\beta^2)~~~(c' = c) 
  \end{equation} 
      However (4.4) remains unchanged:
        \begin{equation}  
     \Delta t'_{{\rm RS}}-  \Delta t'_{{\rm No~ RS}} = \frac{\beta L}{c}+ {\rm O}(\beta^2)~~~(c' = c) 
  \end{equation}
      The prediction $ \tilde{t}'_2- \tilde{t}'_1 = L(1+\beta)/c$ is that of the Sagnac effect in the absence 
of RS (Eq.~(4.3)), or alternatively (Eq.(5.3)) that of conventional SR (RS and the RPVAR). In the latter case the
  experimentally confirmed Sagnac effect is also predicted to not exist, so that the prediction, $c' = c$, of 
   the conventional RPVAR is untenable.
  Experimental verification of the relation (4.3)
  then unambigously demonstrates the non-existence of the RS effect.

 \SECTION{\bf{Relativity of simultaneity test using GPS satellites: Confusion and consternation}}

\begin{figure}[htbp]
\begin{center}\hspace*{-0.5cm}\mbox{
\epsfysize15.0cm\epsffile{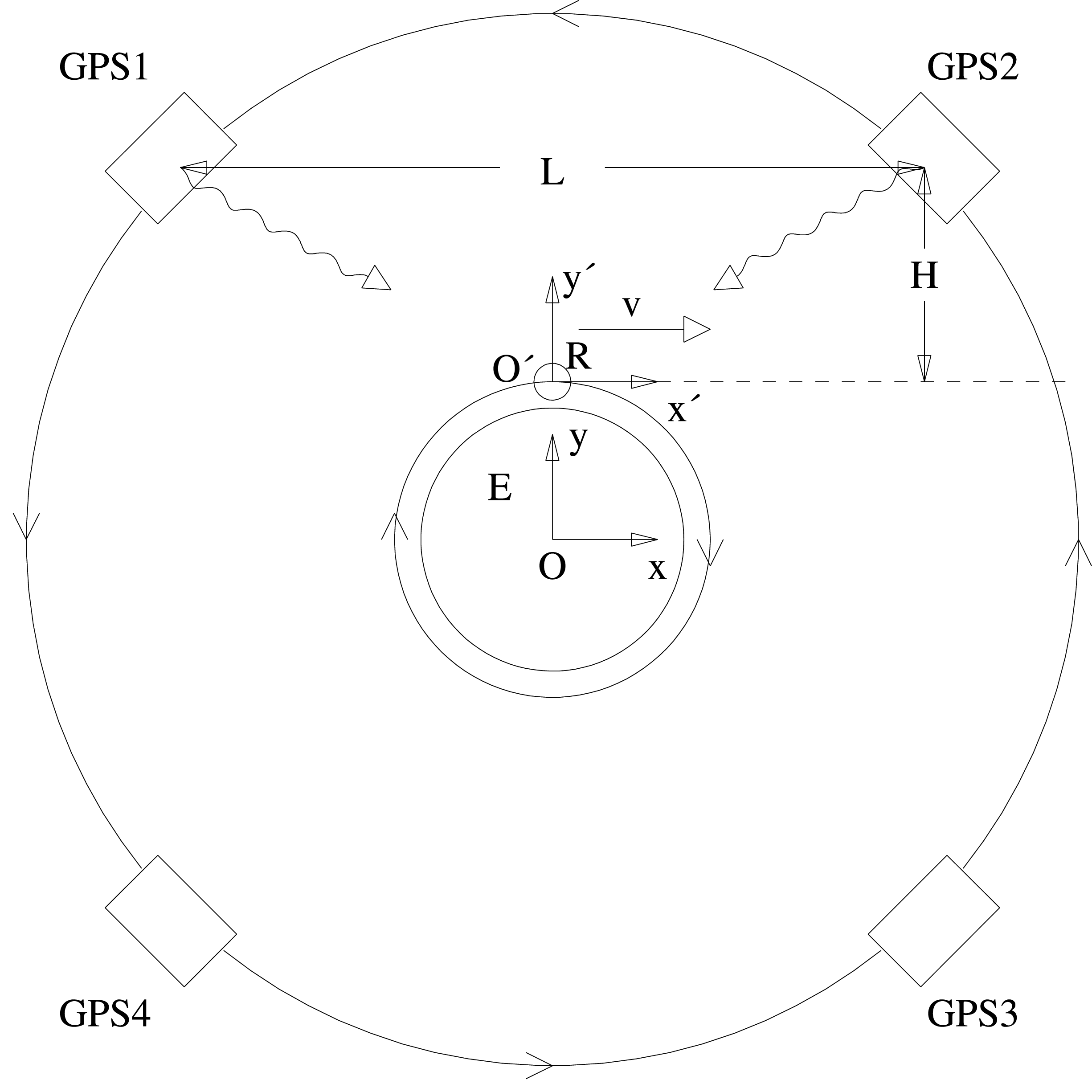}}
\caption{{\em Test of relativity of simultaneity using GPS satellites as transmitters
   and a satellite, R, in low Earth orbit as receiver. The sizes of the Earth, E, and
   the orbit of the GPS satellites are shown approximately to scale; for clarity
    the size of the orbit of R is enlarged. Signals emitted synchronously in the ECI
  frame (S') by GPS1 and GPS2 are received at R when it is at $x = 0$. According to the
  relativity of simultaneity
  of standard special relativity, the signals are received, in the proper frame, S', of R,
  with a 
  time difference of $\Delta t = v L/c^2$, where $v$ is the velocity of R relative to the 
  ECI frame. The figure shows the spatial configuration as seen by an observer in the
   latter frame at the instant that the signals are sent.}}
\label{fig-fig3}
\end{center}
\end{figure}

\begin{figure}[htbp]
\begin{center}\hspace*{-0.5cm}\mbox{
\epsfysize15.0cm\epsffile{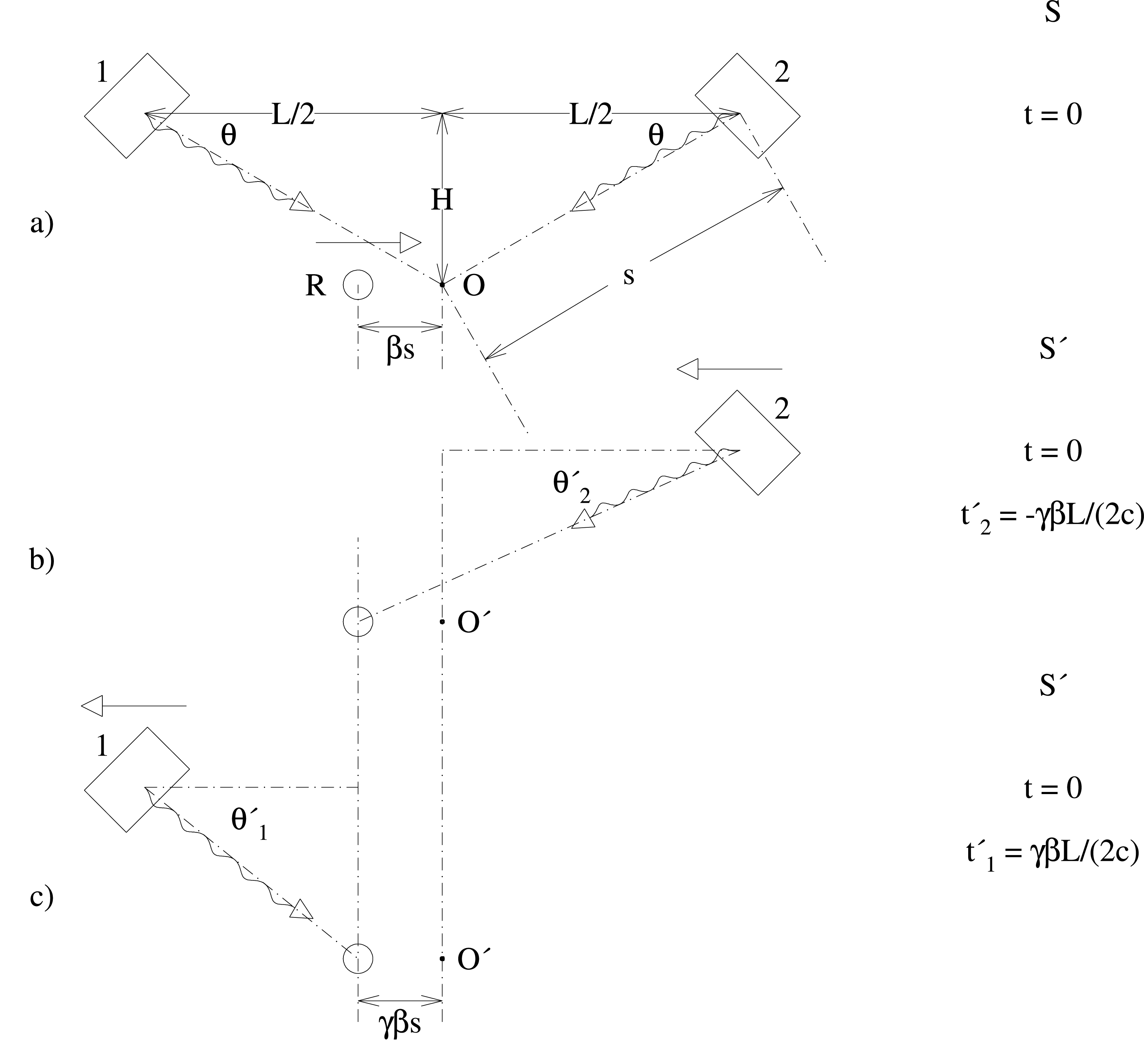}}
\caption{{\em Spatial configurations in different frames at the instant of emission
  of the microwave signals from the satellite GPS1 (1) and GPS2 (2) in the experiment 
  shown in Fig.~3. a) in the ECI frame S,
  b) in the comoving inertial frame S', of the satellite R, at the time of emission of the signal from 1.
  and c) in  S' at the time of emission of the signal from 2.The configurations in b) and c)
  are calculated using Eqs.~(2.1) and (2.2) with $b = 1/a = \gamma$, $d = 1$. See text for
  discussion.}}
\label{fig-fig4}
\end{center}
\end{figure}

   \par An interesting variant of the experiment described in Section 4 above
     uses the synchronised
     satellite-borne clocks of the Global Positioning System (GPS). The latter provide, after
     correction of their proper times
     for the effects of relativistic time dilation and gravitational blue shift, 
     a coordinate time, $t_C$, defined in the ECI reference
     frame~\cite{AshbyPT}. Associating, as before, the ECI frame with S (coordinates ($t$; $x$, $y$, $z$) and the 
     comoving inertial frame of a single satellite, $R$, in low Earth orbit, with the frame S' with coordinates
     ($t'$; $x'$, $y'$, $z'$) the scheme of the proposed experiment is shown in Fig.~3. It is assumed, for
      simplicity, that the satellite's orbit is in the same plane as the orbits of a constellation of four GPS satellites 
    GPS1-GPS4. In the experiment, signals are sent simultaneously, in the ECI frame, from GPS1
    and GPS2, at such 
   a time as to arrive at R when it is in the position, visible from the satellites, at
   which $x = 0$. This corresponds, in GPS nomenclature, to `transmitter time tagging'
  ~\cite{AshbyPT}. The configuration in the frame S at the instant that the signals are sent is shown in more detail
    in Fig.~4a, while configurations in the rest frame, S', of R calculated according to Eqs.~(2.1) and (2.2) 
    for the SR case, at the times when the signals are emitted, in this frame, from GPS2 and GPS1 are shown in Fig.~4b
     and 4c respectively. If $s$ is the flight distance of the microwave signals in the frame S, the satellite
     $R$ is at $x = -\beta s$ in the frame S  at the time of their emission. At the emission
     time of the signals the origins of S and S' are aligned in $x$, and $t=0$. The signal from GPS2 is emitted 
     at time $t'_2 = -\gamma \beta L/(2c)$ at an angle $\theta'_2$ to the $x'$-axis, where
      $\tan\theta'_2 = H/[\gamma(L/2+\beta s)]$ (Fig.~4b). The signal from GPS1 is emitted 
     at the later time, thus manifesting an RS effect,  at time $t'_1 = \gamma \beta L/(2c)$ at
       an angle $\theta'_1$ to the $x'$-axis, where
      $\tan\theta'_2 = H/[\gamma(L/2-\beta s)]$ (Fig.~4c). The flight paths of the signals from
      GPS2 and GPS1 in the receiver rest frame are of length $s'_1$, $s'_2$ respectively.
      If only the leading O($\beta$) terms are retained
     in the formulas, the time dilation and length contraction effects are neglected so that
      the relative velocity of R and the signals is the same in the frames S and S'. If $c'_1$ and $c'_2$ are
     the absolute values, in the frame S', of the velocities of the signals emitted by GPS1 and GPS2 respectively, on 
      making this approximation, then invariance of the $x$ components of the relative velocities of R and the
      signals (the classical Sagnac effect) gives the relations
     \begin{equation}
     c \cos \theta- v = c'_1 \cos\theta'_1,~~~ c \cos \theta+ v = c'_2 \cos\theta'_2
     \end{equation}
        from which follow
    \begin{equation}
      c'_1 = \frac{(c \cos\theta -v)s'_1}{\frac{L}{2}-\beta s},~~ c'_2 = \frac{(c \cos\theta +vc)s'_2}{\frac{L}{2}+\beta s}
    \end{equation}
      where    
   \[ s'_1 = \sqrt{H^2+(L/2-\beta s)^2},~~~ s'_2 = \sqrt{H^2+(L/2+\beta s)^2} \]
     giving for the times-of-flight of the signals in the frame S':
     \begin{equation}
       T'_1 = \frac{s'_1}{c'_1} = \frac{L/2-\beta s}{c \cos\theta -v},
      ~~~ T'_2 = \frac{s'_2}{c'_2} = \frac{L/2+\beta s}{c \cos\theta +v}.
      \end{equation}
      Since, from the geometry of Fig.~4a, $\cos\theta = L/(2s)$, Eqs.~(6.3) give equal times-of-flight
      for the signals:
         \begin{equation}
          T'_1  = T'_1 = \frac{s}{c} = T
          \end{equation}
      where $T$ is the time of flight of either signal in the frame S.
     The difference between the times of reception of the signals at R in the frame S' is then equal to
     the difference between the times of emission, due to the RS effect, in this frame:
       \begin{equation}
        \Delta t' \equiv t'_2-t'_1 =  \frac{\beta L}{c} + {\rm O}(\beta^2)~~~~~~({\rm RS})
         \end{equation}
         which may be compared with $\Delta t'_{{\rm RS}}$ in the experiment described in the previous 
       section as given by Eq.~(4.2). In the case that the event transformation between the frames S and S'
       is performed using Eqs.~(3.1) and (3.2) the RS effect is absent and  $\Delta t' = 0$.
      \par As in the discussion of F\"{u}rth's proposed experiment in the previous section,
 if the Sagnac effect is not correctly taken into account and the speed of the signals is assumed
    to be the same (as predicted by the RPVAR) in the frames S and S', a time-of-flight difference
     equal and opposite to the RS effect $\beta L/c$ is obtained, and again  $\Delta t' = 0$.
       \par The distance parameter $L$ in (6.5) is the separation between two
       adjacent satellites of a GPS constellation: $L = 3.76\times 10^4$ km, to be compared with a typical value
       $L = 400$km in Eq.~(4.2). With $\beta = 2.56\times 10^{-5}$ the value of $\Delta t'$ in Eq.~(6.5)
        is 3.3 $\mu s$ or 960m at the speed of light. This may be compared with the horizontal spatial accuracy
    of 100 m for the Standard Positioning Service (SPS), or of 22 m for the
    Precision Positioning Service (PPS)~\cite{GPSOV}. As for the experiment shown in Fig.1,
    the expected effect is sufficiently large to be observed, with a huge statistical 
    significance, in a single `pass' of the experiment. The possibility to observe such an
    effect, using the GPS, has been previously mentioned in Ref.~\cite{AshbyPT} in a section entitled
    `Confusion and consternation': 
      \begin{quotation}   
      {\sl A 1995 meeting sponsored by the Army Research Laboratory considered the case of 
         a rapidly moving GPS receiver. Did one, in such a case, need a coordinate system with 
         its origin attached to the receiver in order properly to deal with clock synchronisation?
          From the fast-moving receiver's point of view it would seem that the GPS satellite
        clocks would not be synchronised. One can estimate the discrepancies from the approximate 
       synchronisation correction $vx/c^2$, where $v$ is the receiver's speed through the ECI 
      frame and $x$ is the distance to the GPS satellite in question. Suppose the receiver
      is itself in low Earth orbit (7.6 km/s) and the GPS satellite is 20~000 km ahead. Then the
      synchronization  correction  comes to 1.7 $\mu s$. That's enough time for an electromagnetic
      signal to travel 500 m, so one would have to correct for it.
      \par Within the framework of general relativity, however one coordinate system should be as
      legitimate as another. Measurements by an observer travelling with a moving receiver can just
      as well be described in another reference frame by using transformations that relate the two frames.
      In the special case of two inertial frames in uniform relative motion, these are the familiar
      Lorentz transformations.}
     \end{quotation}
       \par Indeed it is just the `familiar Lorentz transformations' that predict the huge time difference
         $\Delta t'$,  found in Eq.~(6.5), corresponding to the ECI frame configuration shown in Fig.~3 and Fig.~4a,
         which is just the one discussed in the passage quoted above from Ref.~\cite{AshbyPT}.
      \par  The effect was incorrectly
    estimated in Ref.~\cite{AshbyPT} by using the shortest distance between a GPS satellite and the low Earth
    orbit, rather than the distance $L$ in Fig.3.
    However, order of magnitude of the effect was correctly given, even though  no definite
    experimental test was proposed and there was no explanation why the very large correction predicted
    is, apparently, not applied in operating the GPS. 
    \par In view of the large observable effect predicted for the experiment in Fig.~3,
    it is interesting to consider a similar experiment where the GPS receiver is not on 
    the satellite R, but at a fixed point on the Earth's surface, as in the usual operational
    mode of the GPS. In this case $\beta$ in the formula $\Delta t = \beta L/c$
    is not the orbital velocity of R of $\simeq 7.6$ km/sec but is rather the projection
    into the plane of the GPS satellite orbits of the velocity of the receiver due to
    the rotation of the Earth. The maximum effect occurs for a receiver at the Equator, viz:
    $v_{rot} \cos 55^{\circ} = 0.47 \times 0.574 = 0.27$ km/sec\footnote{The planes of the
    orbits of GPS satellites are at an angle of $55^{\circ}$ to the equatorial
   plane of the Earth ~\cite{AshbyPT}.}. This corresponds to a value
     of $\Delta t$ of 113 ns, or 34 m at the speed of light. As this is of the same order as the
     PPS accuracy, and three times smaller than the SPS accuracy, no appreciable effects due
     to the non-vanishing of $\Delta t'$, in the case that RS exists, is to be expected on the SPS performance
     at fixed
     points on the surface of the Earth.

 \SECTION{\bf{Summary}}
     As discussed in Ref.~\cite{JHF3}, of the three space-time `effects' of SR: RS, LS and TD, only the
     latter has been, to date, experimentally verified. The RS effect predicts that two simultaneous
      events in a frame S', separated by a distance $L$ in the direction of motion of another
      inertial frame, S, will be observed with a time separation $\delta t = \gamma \beta L/c$ in this last
      frame, where $v = \beta c$ is the relative velocity of the two frames.
    
     \par In the first proposed experiment S' is the co-moving inertial frame of two satellites
      in low  Earth orbit, separated by a distance of a few hundred kilometers, while S is the 
      frame of a microwave transmitter on the surface of the Earth. The value of $\delta t$ for
      the pair of events considered is $\simeq 30$ ns to be compared with the time resolution
      of a similar experiment~\cite{NAVEX} of $\le$ 1 ns. This experiment is in many ways similar
      to that~\cite{Furth} proposed, also employing microwave signals, by F\"{u}rth some forty-six years ago,
       that is discussed in Section 5 above.   
     \par In the second proposed experiment, S is the ECI frame of the GPS satellite
       system, while the GPS signal receiver is in the co-moving inertial
       frame S' of a single satellite in low Earth orbit. The corresponding value of $\delta t$
      is 3.2 $\mu$s as compared to the time resolution of $\simeq 70$ ns for the PPS system~\cite{ISS}.
       Time shifts of this order of magnitude in similar experiments have been previously pointed out~\cite{AshbyPT}
       but no corresponding test for the RS effect proposed.
       For both experiments, all systematic effects in time interval measurements are expected to be
      completely negligible in
       comparison with the predicted values of $\delta t$, in the case that the RS effect exists.
    \par In another paper by the present author~\cite{JHF3} it has been suggested, in 
    order to avoid certain casual paradoxes of SR~\cite{JHF2}, and to ensure translational
    invariance, that the origin of the frame S' in the LT should be, in
    all cases, chosen to coincide with the position of the transformed event (a `local' LT).
    This is equivalent to the use of Eqns(3.1) and (3.2) above with $b = 1/a = \gamma$ to describe
    the clocks. In this case it is predicted that $\delta t = 0$ in both of the experiments presented above,
    i.e. that there is no RS effect

\pagebreak

\end{document}